\documentstyle[floats,aps,epsfig,prb]{revtex}

\begin{document}
\draft

\twocolumn[\hsize\textwidth\columnwidth\hsize\csname
@twocolumnfalse\endcsname

\title{Maximally-localized Wannier functions for entangled energy bands}

\author{Ivo Souza$^{1}$, Nicola Marzari$^{2}$, and David Vanderbilt$^{1}$}

\address{$^{1}$ Department of Physics and Astronomy, Rutgers
                University, Piscataway, New Jersey 08854-8019 \\
         $^{2}$ Department of Chemistry, Princeton University,
                Princeton, New Jersey 08544-1009}

\date{August 4, 2001}
\maketitle

\begin{abstract}
We present a method for obtaining well-localized Wannier-like functions (WFs) 
for energy bands that are attached to or mixed with other bands. The present 
scheme removes the limitation of the usual maximally-localized WFs method 
(N. Marzari and D. Vanderbilt, Phys. Rev. B {\bf 56}, 12847 (1997))
that the bands of interest should form an isolated
group, separated by gaps from higher and lower bands everywhere in the 
Brillouin zone.
An energy window encompassing $N$ bands of interest is specified by the
user, and the algorithm then proceeds to disentangle these from
the remaining bands inside the window by filtering out an optimally 
connected $N$-dimensional subspace. 
This is achieved by minimizing a functional that 
measures the subspace dispersion across the Brillouin zone.
The maximally-localized WFs for the optimal subspace are then obtained
via the algorithm of Marzari and Vanderbilt. The method,
which functions as a postprocessing step using
the output of conventional electronic-structure codes, is 
applied to the $s$ and $d$ bands of copper, and to the valence and low-lying
conduction bands of silicon. For the low-lying nearly-free-electron bands of
copper we find WFs which are centered at the tetrahedral interstitial sites, 
suggesting an alternative tight-binding parametrization.
\end{abstract}

\vskip1pc]

\section{INTRODUCTION}
\label{sec:intro}

When studying electrons in solids, it is often the case that
only a small subset of the available one-electron states
contributes significantly to the properties under consideration.
Moreover, the states of
interest typically lie within a limited energy range.
For instance, for modeling electron transport or magnetic properties,
only the partially filled bands
close to the Fermi energy $E_F$ are needed.
This is the rationale behind the tight-binding and Hubbard models,
in which only a few energy bands are kept.\cite{am,fradkin}
Those models rely on the existence of a minimal set of
spatially localized orbitals spanning the manifold of relevant states.

In recent years there has been growing interest in explicitly
constructing such orbitals from first-principles density-functional
calculations. One potential application consists in
obtaining the parameters in correlated Hamiltonians by constraining the 
occupation of the orbitals to find
the energy cost of deviating from the mean-field solution  (``constrained 
density-functional theory''\cite{hybertsen89,mcmahan90}). 
Another arises in the context of the ``dynamical mean-field theory''
which, when combined with density-functional methods, requires the 
specification of localized
orbitals describing the narrow bands of interest.\cite{anisimov97}

Wannier functions\cite{wannier37} (WFs) are 
a very natural type of localized 
orbital for extended systems.
They play a central role in formal discussions of the tight-binding\cite{am}
and Hubbard\cite{fradkin} models. Traditionally they have often been
invoked --~although rarely calculated explicitly~--  as a
convenient basis for describing local phenomena, such as impurities,\cite{yu} 
excitons,\cite{yu} and magnetic properties.\cite{slater37}
More recently, WFs have found important applications in connection with 
linear-scaling
algorithms for electronic structure calculations.\cite{goedecker99}
Moreover, they play an important role in
the theory of electronic polarization and localization in insulators, with the
former quantity being related to the centers of charge of the 
WFs\cite{ksv93,nunes94} 
and the latter to their quadratic spreads.\cite{resta99,souza00} 
These developments have also led to 
generalizations of the concept of Wannier functions to correlated electron 
systems.\cite{souza00,kudinov99,koch01}

The main obstacles to the construction of WFs
in practical calculations have been their nonuniqueness 
(or ``gauge dependence'') and the difficulties in dealing with
degeneracies among the Bloch states. These have been overcome by the
development by Marzari and Vanderbilt
 of a general and practical method for extracting
``maximally-localized'' WFs from an isolated group of bands.\cite{mv97}
(By ``isolated'' we mean a group of bands that may become entangled with 
one another across the Brillouin zone, but is separated from all other bands by
finite gaps throughout the entire Brillouin zone. The set of valence bands of
an insulator constitutes an important example.) 
The method has been
successfully used to describe the dielectric properties of several insulating 
systems, such as crystalline\cite{mv97} and amorphous\cite{a-si}
 semiconductors, ferroelectric perovskites,\cite{mv98}
liquid water,\cite{silvestrelli99} compressed solid hydrogen,\cite{souza00b}
and manganese oxide.\cite{posternak01}
It has been implemented for plane-wave,\cite{mv97} linear augmented
plane-wave,\cite{posternak01} and tight-binding\cite{souza00b} basis sets.

However, in many cases
the group of bands of interest is not isolated in the above sense, especially 
when dealing with metals or with the
empty bands of insulators. For example,
the conduction $s$ band of an alkali metal is attached at points or lines of
high symmetry to higher bands; the $d$ bands of a noble or transition metal
are hybridized with an $s$ band, which in turn is attached to higher bands;
the conduction bands of a copper-oxide superconductor emerge from 
a dense group of bands below; and the four low-lying antibonding bands of a
tetrahedral semiconductor are connected to higher conduction bands. 

A successful technique that has been applied for constructing localized
orbitals that describe such bands is the 
``downfolding'' technique\cite{andersen95,andersen00} that has been developed
for electronic structure methods based on muffin-tin orbitals. 
There have also been previous attempts at constructing WFs for
non-isolated groups of bands, namely for noble and transition 
metals\cite{goodings69,modrak87,modrak92,sporkmann94} and for
tetrahedral semiconductors.\cite{teichler71,sporkmann97} 
These attempts fall into two categories: (i) the WFs are obtained
directly from a variational principle, as suggested by
Kohn,\cite{kohn73} or (ii) they are obtained as Fourier transforms
of Bloch functions, with the help of a model Hamiltonian that reproduces the
band structure in the desired energy range, as suggested by
Bross.\cite{bross71}

We will describe an alternative Wannier-based approach that is closer in spirit
to the Fourier transform method of Bross and co-workers, but does not require 
the construction of an auxiliary model Hamiltonian. The method can be regarded
as an extension to the case of attached bands of the maximally-localized WF 
method of Marzari and Vanderbilt.\cite{mv97} It has the desirable features that
it can be implemented with any basis set (e.g., plane waves),
and requires minimal user-intervention (the only ``adjustable parameter''
being a specification of the
energy range of interest). Like the approach of Ref.~\onlinecite{mv97}, ours 
is a ``postprocessing'' method, taking as its input the Bloch eigenstates and
eigenvalues calculated by a standard electronic-structure code.

Strictly speaking, the resulting orbitals are not WFs (or even ``generalized 
WFs''\cite{mv97}) in
the usual sense. They are nevertheless Wannier-like in the fundamental sense 
that they are obtained via an integral over the Brillouin zone of 
Bloch-like functions.
As such they form an orthonormal, localized basis of the same Bloch
subspace from which they were constructed.

The power of the present approach is illustrated by one particularly
striking result that emerged from the work.  In Sec.~\ref{sec:copper_blobs}
we find that a rather natural representation of the low-lying
bands of an fcc metal 
like copper can be made in terms of a set of
five Cu $d$-like WFs and two additional WFs centered at the
tetrahedral interstitial locations.  This provides a basis for a
concise tight-binding representation of copper that has not, to our
knowledge, previously been considered.

The paper is organized as follows. In Sec.~\ref{sec:maxloc} we review the 
method of Marzari and Vanderbilt
for obtaining well-localized WFs for an isolated group
of bands. In Sec.~\ref{sec:dwf} we describe our procedure for dealing
with attached energy bands, and in Sec.~\ref{sec:results} we illustrate it with
a set of applications. 
Finally, in Sec.~\ref{sec:conclusions} we present a summary and conclusions.

\section{Maximally-localized Wannier functions for an isolated group of bands}
\label{sec:maxloc}

A set of WFs $w_{n{\bf R}}({\bf r})=w_n({\bf r}-{\bf R})$ labeled by 
Bravais lattice vectors ${\bf R}$ can be constructed from the
Bloch eigenstates $\psi_{n \bf k}$ of band $n$ using the unitary 
transformation
\begin{equation}
\label{wf}
w_{n{\bf R}}({\bf r}) = \frac{v}{8 \pi^3} \int_{\rm BZ}
e^{- i {\bf k} \cdot {\bf R}} \, \psi_{n \bf k} \, d{\bf k},
\end{equation}
where $v$ is the volume of the unit cell of the crystal and
the integral is over the Brillouin zone.
Except for the constraint $\psi_{n,{\bf k}+{\bf G}} = \psi_{n \bf k}$
for all reciprocal lattice vectors ${\bf G}$,
the overall phases of the Bloch functions
$\psi_{n \bf k} = e^{i {\bf k} \cdot {\bf r}} u_{n \bf k}$ are at our 
disposal. However, a different choice of phases (or ``gauge''),
\begin{equation}
\label{gauget_isol}
u_{n \bf k} \rightarrow e^{i \varphi_n({\bf k})} \, u_{n \bf k},
\end{equation}
does not translate into a simple change of the overall phases of the
WFs; their shape and spatial extent will in general be affected.
If the band is isolated, Eq.~(\ref{gauget_isol}) is the only allowed type of
gauge transformation for changing the set of WFs 
$w_n({\bf r}-{\bf R})$ associated with that 
band. In the case of an isolated group of $N$ bands, the allowed
transformations are of the more general form
\begin{equation}
\label{gauget_composite}
u_{n \bf k} \rightarrow \sum_{m=1}^N \, U_{mn}^{(\bf k)} \, u_{m \bf k},
\end{equation}
where $U^{(\bf k)}$ is a unitary matrix that mixes the bands at
wave vector ${\bf k}$. The resulting orbitals are called
``generalized Wannier functions''.\cite{mv97}

Once a measure of localization has been chosen and an isolated group of bands
specified, the search for the corresponding set of 
``maximally-localized'' WFs becomes a problem of functional minimization
in the space of the matrices $U^{(\bf k)}$.
The strategy of Ref.~\onlinecite{mv97} consists in minimizing
the sum of the quadratic spreads of the Wannier probability distributions
${| w_n({\bf r}) |}^2$,
\begin{equation}
\label{omega}
\Omega = \sum_{n=1}^N \, \left( {\left< r^2 \right>}_n - 
{\left< {\bf r} \right>}_n^2 \right),
\end{equation}
where the sum is over the chosen group of bands and
${\left< {\bf r} \right>}_n = \int \, {\bf r} \, {| w_n({\bf r}) |}^2 \, 
d{\bf r}$, etc. Interestingly,
the resulting ``maximally-localized'' (or ``maxloc'') WFs turn out to be
real, apart from an arbitrary overall phase factor. 

In numerical calculations 
the Bloch states $\psi_{n \bf k}$ are computed on a regular
mesh of $k$-points in the Brillouin zone;
the integral in Eq.~(\ref{wf}) is then replaced by a sum over the points
in the mesh. In Ref.~\onlinecite{mv97} an expression was derived for the 
gradient of the spread functional $\Omega$ with respect to an
infinitesimal rotation $\delta U^{({\bf k})}$ of the set of Bloch 
orbitals.
The only information needed for calculating the gradient
are the overlaps 
\begin{equation}
\label{overlap}
M_{mn}^{({\bf k},{\bf b})}=
\left< u_{m \bf k} | u_{n,{\bf k}+{\bf b}} \right>,
\end{equation}
where ${\bf b}$ are vectors connecting a mesh point to its
near neighbors. Once the gradient is computed, the minimization can
proceed via a steepest-descent or conjugate-gradients algorithm.

In Ref.~\onlinecite{mv97} the spread $\Omega$ was decomposed into two terms,
\begin{equation}
\label{omega_sum}
\Omega = \Omega_{\rm I} + \widetilde{\Omega},
\end{equation}
both of them non-negative.  The first measures the $k$-space dispersion
of the band projection operator, while the second reflects the extent
to which the Wannier functions fail to be eigenfunctions of the
band-projected position operators.
$\Omega_{\rm I}$ will play a central role
in the present work. For an isolated group
of bands it is invariant under any gauge
transformation~(\ref{gauget_composite}), so that 
minimizing $\Omega$ amounts to minimizing $\widetilde{\Omega}$.
When using a regular mesh of $k$-points, $\Omega_{\rm I}$ is given by
\begin{equation}
\label{omega_i}
\Omega_{\rm I}= \frac{1}{N_{\rm kp}} \,
\sum_{{\bf k},{\bf b}} \, w_b \,
\sum_{m=1}^N 
\Big[ \,
  1 - \sum_{n=1}^N \,
  {\big| M_{mn}^{({\bf k},{\bf b})} \big|}^2 \,
\Big],
\end{equation}
where $N_{\rm kp}$ is the total number of $k$-points, $N$ is the 
number of bands in the group, and $w_b$ is a weight that arises from the
discretization procedure by which derivatives with respect to ${\bf k}$ are
approximated by finite differences.\cite{mv97} The corresponding
expression for $\widetilde\Omega$ can be found in Ref.~\onlinecite{mv97}.

\section{Maximally-localized Wannier functions for attached bands}
\label{sec:dwf}

\subsection{Description of the method}
\label{sec:description}

\begin{figure}
\centerline{\epsfig{file=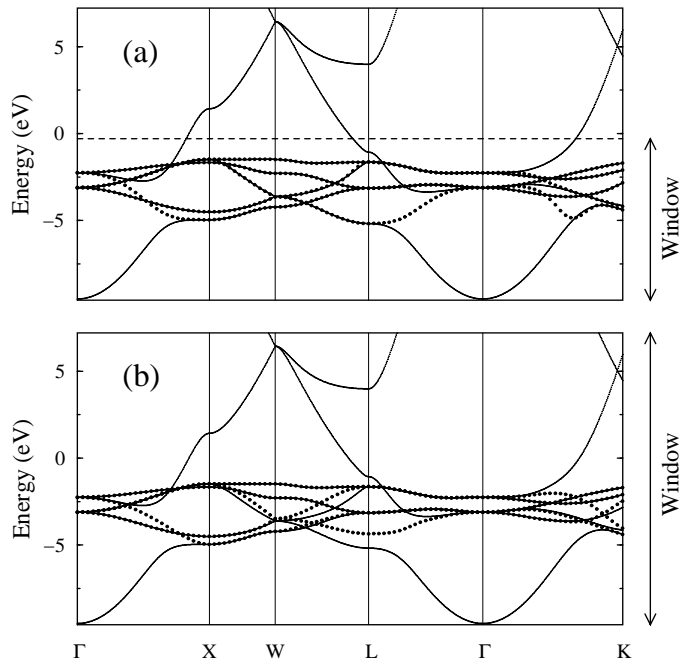,width=3.4in,angle=-90}}
\vspace{0.2cm}
\caption{
Solid line: Calculated band structure of copper.
Dotted line: Interpolated bands obtained from the five
$d$-like Wannier functions. (a) and (b) differ in the
choice of the energy window used to compute the Wannier functions
($[-9.59,-0.29]$~eV in (a) and $[-9.59,7.21]$~eV in (b)). The
zero of the energy scale is at the Fermi energy.
}
\label{fig:Cu_d_bands}
\end{figure}

For definiteness let us suppose we want to
``disentangle'' the five $d$ bands of copper from the $s$ band which crosses 
them (see Fig.~\ref{fig:Cu_d_bands}) and construct a set of well-localized
WFs associated with the resulting $d$ bands.
Heuristically the $d$ bands are the five narrow bands and the
$s$ band is the wide band. 
The difficulty arises because there are regions of $k$-space where all
six bands are close together, so that as a result of hybridization
``the distinction between $d$-band and $s$-band levels is not meaningful''
(Ref.~\onlinecite{am}, p.~288).

Let us now outline our strategy, which can be divided in two steps. First we 
cut out an energy window that encompasses the $N$ bands of interest
($N=5$ in our example). Figs.~\ref{fig:Cu_d_bands}(a) and
\ref{fig:Cu_d_bands}(b) correspond
to different choices for this
energy window. At each $k$-point the number $N_{\bf k}$ of bands
that fall inside the window is equal to or larger than the target number
of bands $N$. This procedure defines an $N_{\bf k}$-dimensional
Hilbert space ${\cal F}({\bf k})$ spanned by the states
$u_{n \bf k}$ within the window. If at some ${\bf k}$
$N_{\bf k}=N$, there is nothing to do there; if $N_{\bf k}>N$ our aim is to
find the $N$-dimensional subspace 
${\cal S}({\bf k}) \subseteq {\cal F}({\bf k})$ that, among all possible 
$N$-dimensional subspaces of ${\cal F}({\bf k})$, leads to the smallest 
$\Omega_{\rm I}$ (Eq.~(\ref{omega_i})).
(Recall that for an isolated group of bands $\Omega_{\rm I}$ is
gauge-invariant, since it is an intrinsic property of the manifold of states.
Thus $\Omega_{\rm I}$ can be regarded as a functional of
${\cal S}({\bf k})$.)
In the second step we work within
the optimal $N$-dimensional subspaces ${\cal S}({\bf k})$ 
selected in the first step, and minimize $\widetilde{\Omega}$ using the
algorithm of Marzari and Vanderbilt\cite{mv97} summarized in the previous
section. The end result is a set of $N$ maximally localized WFs and the 
corresponding $N$ energy bands. We emphasize that it is the 
first step (minimization of $\Omega_{\rm I}$) that is new with respect to 
Ref.~\onlinecite{mv97}.

\subsection{Physical interpretation of $\Omega_{\rm I}$}
\label{sec:omega_i}

Why is minimizing $\Omega_{\rm I}$ a sensible strategy for picking
out the $d$-bands? This can be understood by noting that
heuristically $\Omega_{\rm I}$
measures the ``change of character'' of the states across the Brillouin 
zone.\cite{mv97} Indeed, Eqs.~(\ref{overlap}) and (\ref{omega_i}) show that
$\Omega_{\rm I}$ is small 
whenever ${| \left< u_{n \bf k} | u_{m,{\bf k}+{\bf b}} \right>|}^2$, 
the square of
the magnitude of the overlap  between states at nearby $k$-points, is 
large. Thus by minimizing $\Omega_{\rm I}$ we are choosing 
self-consistently
at every ${\bf k}$ the subspace ${\cal S}({\bf k})$ that has minimum
``spillage'' or mismatch (see below) as ${\bf k}$ is varied. In the
present example this optimal ``global smoothness of connection'' will be 
achieved by keeping the five 
well-localized 
$d$-like states
and excluding the 
more delocalized 
$s$-like state.
We will gain more intuition about the meaning of minimizing $\Omega_{\rm I}$ 
while discussing specific examples in Sec.~\ref{sec:results}.

What is meant by ``spillage''\cite{mv97,daniel95} becomes clear once we rewrite
Eq.~(\ref{omega_i}) as
\begin{equation}
\label{omega_i_spillage}
\Omega_{\rm I}=\frac{1}{N_{\rm kp}}\sum_{{\bf k},{\bf b}} \,
w_b \, T_{{\bf k},{\bf b}}
\end{equation}
with
\begin{equation}
\label{spillage}
T_{{\bf k},{\bf b}}=N-\sum_{m,n} \, {|M_{mn}^{({\bf k},{\bf b})}|}^2
= {\rm tr} [\hat{P}_{\bf k}\,\hat{Q}_{{\bf k}+{\bf b}}],
\end{equation}
where 
$\hat{P}_{\bf k}=\sum_n \, | u_{n \bf k} \rangle 
\langle u_{n \bf k} |$ 
is the projector onto ${\cal S}({\bf k})$,
$\hat{Q}_{\bf k}={\bf 1}-\hat{P}_{\bf k}$, and the band indices $m,n$ run over
$1, \ldots, N$. $T_{{\bf k},{\bf b}}$ is called the ``spillage'' between the 
spaces ${\cal S}({\bf k})$ and ${\cal S}({\bf k}+{\bf b})$ because it measures
the degree of mismatch between them, vanishing when they are identical.

Further discussion of the geometrical and physical interpretation of
$\Omega_{\rm I}$ can be found in Refs.~\onlinecite{souza00} and 
\onlinecite{mv97}.
In particular, it has been shown that the value of $\Omega_{\rm I}$ associated
with the valence bands of an insulator is the experimentally measurable 
mean-square quantum
fluctuation of the ground state macroscopic polarization.\cite{souza00}
This can be interpreted as the quadratic spread of an appropriately defined
collective center-of-mass distribution for the electrons, and can be recast
as an electronic localization length squared. Hence  our procedure of 
minimizing $\Omega_{\rm I}$ selects the $N$-dimensional subspaces
${\cal S}({\bf k})$ where the electrons
are most localized in the above sense (assuming for the purpose of this 
argument that all the electron states in those subspaces are occupied).

Finally we note in passing that our two-step procedure of minimizing first
$\Omega_{\rm I}$ and then $\widetilde{\Omega}$ is in principle different
from directly minimizing their sum $\Omega$. In view of the discussion
presented above, we believe that the procedure adopted here is conceptually the
more natural of the two, although we would expect them
to yield similar results in practice.
Also, as we will now show, the separate minimization
of $\Omega_{\rm I}$ turns out to be a particularly simple and robust
procedure.

\subsection{Iterative minimization of $\Omega_{\rm I}$}
\label{sec:min}

\begin{figure}
\centerline{\epsfig{file=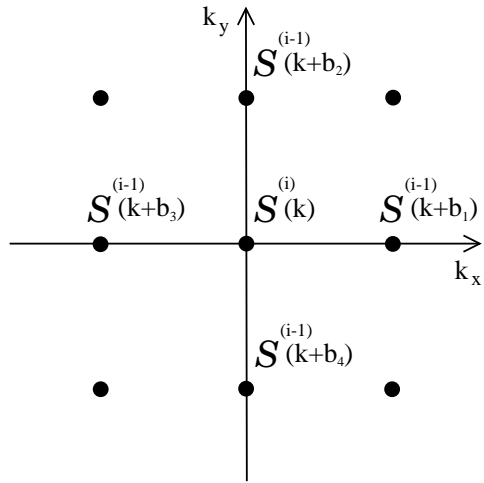,width=2.5in}}
\vspace{0.2cm}
\caption{
Schematic representation of the subspaces of Bloch-like states on a
grid of $k$-points. Our procedure consists of iteratively minimizing the
``spillage'', or degree of mismatch (see text), between the subspaces at
neighboring $k$-points.
}
\label{fig:k_grid}
\end{figure}

Since the functional~(\ref{omega_i}) that we wish to minimize couples states at
different $k$-points, the problem has to be solved self-consistently 
throughout the Brillouin zone. 
Our strategy  is to proceed iteratively until the optimal
``global smoothness of connection'' is achieved. On the $i$-th iteration we 
go through all the $k$-points in the grid, 
and for each of them we find $N$ orthonormal
states $u_{n \bf k}^{(i)}$, defining a subspace
${\cal S}^{(i)}({\bf k}) \subseteq {\cal F}({\bf k})$ such that the 
``spillage'' over the neighboring subspaces ${\cal S}^{(i-1)}({\bf k}+{\bf b})$
from the previous iteration is as small as possible 
(Fig.~\ref{fig:k_grid}). 

Using Lagrange multipliers to enforce orthonormality, the stationarity 
condition at the $i$-th iteration reads
\begin{equation}
\label{lagrange_mult}
\frac{\delta \Omega_{\rm I}^{(i)}}{\delta u_{m \bf k}^{(i)*}}
+\sum_{n=1}^N \, \Lambda_{nm,\bf k}^{(i)} \, \frac{\delta}
                            {\delta u_{m \bf k}^{(i)*}}
\left[
  \left< \left. u_{m \bf k}^{(i)} \right| u_{n \bf k}^{(i)} 
\right>-\delta_{m,n}
\right]=0,
\end{equation}
where $\Lambda_{\bf k}^{(i)}$ is an $N \times N$ matrix. Let
\begin{equation}\label{omega_i_b}
\Omega_{\rm I}^{(i)} = \frac{1}{N_{\rm kp}} \sum_{{\bf k}=1}^{N_{\rm kp}}
\omega_{\rm I}^{(i)}({\bf k})
\end{equation}
where, according to Eq.~(\ref{omega_i_spillage}),
\begin{eqnarray}
\label{omega_i_k}
\omega_{\rm I}^{(i)}({\bf k})&=&
\sum_{\bf b} \, w_b \, T_{{\bf k},{\bf b}}^{(i)} \nonumber\\ 
&=&\sum_{\bf b} \, w_b \,
\sum_{m=1}^N 
\Big[ 
  1 - \sum_{n=1}^N 
  {\left|
  \left< \left. u_{m \bf k}^{(i)} \right| u_{n,{\bf k}+{\bf b}}^{(i-1)} \right>
  \right|}^2
\Big].
\end{eqnarray}
The first term in Eq.~(\ref{lagrange_mult}) now becomes
\begin{equation}
\label{var_omega_i}
\frac{\delta \Omega_{\rm I}^{(i)}}{\delta u_{m \bf k}^{(i)*}}
=\frac{1}{N_{\rm kp}}
\left\{
  \frac{\delta \omega_{\rm I}^{(i)}({\bf k})}
       {\delta u_{m \bf k}^{(i)*}} \, + \,
  \sum_{\bf b} \,
  \frac{\delta \omega_{\rm I}^{(i)}({\bf k}+{\bf b})}
       {\delta u_{m \bf k}^{(i)*}}
\right\}.
\end{equation}
From Eq.~(\ref{omega_i_k}) we find
\begin{equation}
\label{var_omega_i_k}
\frac{\delta \omega_{\rm I}^{(i)}({\bf k})}
     {\delta u_{m \bf k}^{(i)*}}=
-\sum_{\bf b} \, w_b \, \hat{P}_{{\bf k}+{\bf b}}^{(i-1)}
\left| \left. u_{m \bf k}^{(i)} \right> \right.,
\end{equation}
where $\hat{P}_{{\bf k}+{\bf b}}^{(i-1)}$ is the projector onto
${\cal S}^{(i-1)}({\bf k}+{\bf b})$. Likewise, one easily obtains
\begin{equation}
\label{var_omega_i_k+b}
\frac{\delta \omega_{\rm I}^{(i)}({\bf k}+{\bf b})}
     {\delta u_{m \bf k}^{(i)*}}=
-w_b \, \hat{P}_{{\bf k}+{\bf b}}^{(i-1)} 
\left| \left. u_{m \bf k}^{(i)} \right> \right..
\end{equation}
Combining the previous equations, the stationarity 
condition~(\ref{lagrange_mult}) becomes
\begin{equation}
\label{stationary_cond_a}
\Big[
  \sum_{\bf b} \, w_b \, \hat{P}_{{\bf k}+{\bf b}}^{(i-1)}
\Big]
\left| \left. u_{m{\bf k}}^{(i)} \right> \right.=
\sum_{n=1}^N \widetilde{\Lambda}_{nm,\bf k}^{(i)} 
\left| \left. u_{n{\bf k}}^{(i)} \right> \right.,
\end{equation}
where 
$\widetilde{\Lambda}_{nm,\bf k}^{(i)}=(N_{\rm kp}/2) 
\Lambda_{nm,\bf k}^{(i)}$.
By choosing a unitary transformation that diagonalizes 
$\widetilde{\Lambda}_{\bf k}^{(i)}$, this can be recast as an eigenvalue 
equation:
\begin{equation}
\label{stationary_cond}
\Big[
  \sum_{\bf b} \, w_b \, \hat{P}_{{\bf k}+{\bf b}}^{(i-1)}
\Big]
\left| \left. u_{m{\bf k}}^{(i)} \right> \right.=
\lambda_{m \bf k}^{(i)} \left| \left. u_{m{\bf k}}^{(i)} \right> \right..
\end{equation}
The eigenvalues of the above equation obey
$0 \leq \lambda_{m \bf k}^{(i)} \leq \sum_{\bf b} \, w_b$;
in particular, $\lambda_{m \bf k}^{(i)} < \sum_{\bf b} \, w_b$ 
whenever the eigenstate $u_{m \bf k}^{(i)}$ does not 
lie completely within all of the nearby subspaces 
${\cal S}^{(i-1)}({\bf k}+{\bf b})$. 
Combining Eqs.~(\ref{omega_i_k}) and (\ref{stationary_cond}), we 
find
\begin{equation}
\label{omega_i_k_b}
\omega_{\rm I}^{(i)}({\bf k})=N \, \sum_{\bf b} \, w_b \, - \,
\sum_{m=1}^N \, \lambda_{m \bf k}^{(i)}.
\end{equation}
It is clear from Eqs.~(\ref{omega_i_b}) and (\ref{omega_i_k_b}) that when 
constructing ${\cal S}^{(i)}({\bf k})$ one should pick the $N$ eigenvectors of
Eq.~(\ref{stationary_cond}) with largest eigenvalues, so as to ensure that the 
stationary point corresponds to the absolute minimum of $\Omega_{\rm I}^{(i)}$.

Self-consistency is achieved when 
${\cal S}^{(i)}({\bf k})={\cal S}^{(i-1)}({\bf k})$ at all the grid points. 
We have encountered cases where the iterative procedure 
outlined above was not stable. In those cases, the
problem was solved by using as the input for the present step
a linear mixing of the input and output subspaces from the previous step. 
More precisely, the eigenvalue equation~(\ref{stationary_cond}) was replaced 
by
\begin{equation}
\label{iterative_b}
\Big\{
  \sum_{\bf b} w_b 
  {\left[ \hat{\cal P}^{(i)}_{{\bf k}+{\bf b}} \right]}_{\rm in}
\Big\}
\left| \left. u_{m{\bf k}}^{(i)} \right> \right.=
\lambda_{m \bf k}^{(i)} \left| \left. u_{m{\bf k}}^{(i)} \right> \right.,
\end{equation}
where
\begin{equation}
\label{linear_mix}
{\left[ \hat{\cal P}^{(i)}_{{\bf k}+{\bf b}} \right]}_{\rm in} =
\alpha \hat{P}_{{\bf k}+{\bf b}}^{(i-1)}
+(1-\alpha) {\left[ \hat{\cal P}^{(i-1)}_{{\bf k}+{\bf b}} \right]}_{\rm in}
\end{equation}
with $0 < \alpha \leq 1$.\cite{foot:omega_i_mix}
A typical value is $\alpha$=0.5.

In practice we solve Eq.~(\ref{iterative_b}) in the basis of the original
$N_{\bf k}$ Bloch eigenstates $u_{n \bf k}$ inside the energy window. 
Each iteration then amounts to
diagonalizing the following $N_{\bf k} \times N_{\bf k}$ Hermitian matrix at 
every ${\bf k}$:
\begin{equation}
\label{z_matrix}
Z_{mn}^{(i)}({\bf k})=
\Big< u_{m \bf k}
\Big| \sum_{\bf b} w_b 
  {\left[ \hat{\cal P}^{(i)}_{{\bf k}+{\bf b}} \right]}_{\rm in}
\Big| u_{n \bf k} \Big>.
\end{equation}
Since these are small matrices,
each step of the iterative procedure is computationally cheap.
In particular, the time-consuming computation of the overlap matrices 
$M^{({\bf k},{\bf b})}$ of Eq.~(\ref{overlap}) can be done once and for all
at the beginning, using the original Bloch eigenstates inside the energy
window; their subsequent update during the iterative minimization is
very inexpensive. An analogous situation occurs when updating the matrices
$U^{({\bf k})}$ in Eq.~(\ref{gauget_composite}) during
the minimization of
$\widetilde{\Omega}$ to obtain the ``maxloc'' WFs.\cite{mv97}

\subsection{Initial guess for the subspaces}
\label{sec:initial_guess}

In order to start the iterative minimization of $\Omega_{\rm I}$, the user
should provide an initial guess for the subspaces ${\cal S}({\bf k})$. We have
found that the minimization procedure is quite robust, in the sense that it 
is able to arrive at the global minimum starting from a very rough initial 
guess. In practice we usually select the initial subspaces following a strategy
very similar to the one outlined in Ref.~\onlinecite{mv97} for
starting the minimization of $\widetilde{\Omega}$. 

A set of $N$ localized trial orbitals $g_n({\bf r})$ is chosen corresponding
to some rough initial guess at the WFs, and these are
then projected onto the $N_{\bf k}$
Bloch eigenstates inside the energy window,
\begin{equation}
\label{trail_proj}
\left| \left. \phi_{n \bf k} \right> \right. = \sum_{m=1}^{N_{\bf k}}
A_{mn} \left| \left. \psi_{m \bf k} \right> \right.,
\end{equation}
where 
$A_{mn}= \left< \psi_{m \bf k} \left| g_n \right. \right>$ is an
$N_{\bf k} \times N$ matrix. The resulting $N$ orbitals are then 
orthonormalized via L\"owdin's symmetric orthogonalization 
procedure,\cite{lowdin49} i.e.,
\begin{eqnarray}
\label{orthonormalization}
\left| \left. \psi_{n \bf k}^{(0)} \right> \right. &=& 
\sum_{m=1}^N {(S^{-1/2})}_{mn} \left| \left. \phi_{m \bf k} \right> 
\right. \nonumber \\ 
&=& \sum_{m=1}^{N_{\bf k}} {(A S^{-1/2})}_{mn} 
\left| \left. \psi_{m \bf k} \right> \right.,
\end{eqnarray}
where 
$S_{mn} = \left< \phi_{m \bf k} \left| \phi_{n \bf k} \right. \right>=
{(A^{\dagger} A)}_{mn}$. Finally these Bloch-like functions are converted to 
cell-periodic functions
$u_{n \bf k}^{(0)} = e^{-i {\bf k} \cdot {\bf r}}
\psi_{n \bf k}^{(0)}$.
The matrix $A S^{-1/2}$ can easily be computed by performing the
singular-value decomposition $A=ZDV$,\cite{nrecipes} where
$Z$ and $V$ are $N_{\bf k} \times {N_{\bf k}}$ and
$N \times N$ unitary matrices respectively, and $D$ is $N_{\bf k} \times N$ and
diagonal. This leads to $A S^{-1/2}=Z {\bf 1} V$, where ${\bf 1}$ is the 
$N_{\bf k} \times N$ identity matrix.

\subsection{Minimization of $\widetilde{\Omega}$}
\label{sec:min_omegai}

At the end of the first step of our procedure (minimization of 
$\Omega_{\rm I}$) we are left at each $k$-point with an $N$-dimensional
subspace ${\cal S}({\bf k})$, and for definiteness we diagonalize
the Hamiltonian inside this subspace to obtain
$N$ Bloch-like eigenfunctions $\widetilde{\psi}_{n \bf k}
=e^{i{\bf k}\cdot{\bf r}} \widetilde{u}_{n \bf k}$ 
and eigenvalues $\widetilde{\epsilon}_{n \bf k}$.
The second step is to
find the $N \times N$ unitary matrices $U^{({\bf k})}$ 
(Eq.~(\ref{gauget_composite})) that, applied
to the $\widetilde{\psi}_{n \bf k}$, produce the rotated set of Bloch-like
states that is transformed via~(\ref{wf}) into the maximally-localized WFs 
$w_{n{\bf R}}$. 
This is done using the method of Marzari and Vanderbilt\cite{mv97} for
minimizing $\widetilde{\Omega}$, briefly discussed in Sec.~\ref{sec:maxloc}.
An initial guess for the unitary matrices $U^{({\bf k})}$ is obtained by
projecting a set of $N$ localized orbitals onto the states
$\widetilde{\psi}_{n \bf k}$.
Typically the same set of orbitals is used as in the initialization step for
the minimization of $\Omega_{\rm I}$.
(In our experience, when a particularly bad choice of trial orbitals is made,
the minimization of
$\Omega_{\rm I}$ is less likely to become trapped in local minima than the
minimization of $\widetilde{\Omega}$.)

\subsection{Interpolated band structure}
\label{sec:interpol_bands}

Starting from the ``maxloc'' WFs, the corresponding energy bands can be 
computed at arbitrary points in the Brillouin zone using a Slater-Koster 
interpolation scheme.\cite{sporkmann94,bross71,slater54}
Of course, the interpolation could proceed directly from the non-rotated states
$\widetilde{u}_{n \bf k}$; however, use of the optimally rotated ones
ensures that the interpolated band 
structure is as smooth as possible.\cite{foot:one_band}

The interpolation procedure involves first calculating the
Hamiltonian matrix for the rotated states,
\begin{equation}
\label{ham_rot}
H^{({\rm rot})}({\bf k})= {(U^{({\bf k})})}^{\dagger} \, 
\widetilde{H}({\bf k}) \, U^{({\bf k})},
\end{equation}
where $\widetilde{H}_{mn}({\bf k}) = \widetilde{\epsilon}_{m \bf k}
\delta_{m,n}$.
Next we Fourier transform $H^{({\rm rot})}({\bf k})$ into a set of 
$N_{\rm kp}$ Bravais lattice vectors ${\bf R}$ within a Wigner-Seitz supercell
centered around ${\bf R}=0$:
\begin{eqnarray}
\label{ftt_1}
H_{mn}^{({\rm rot})}({\bf R}) &=& 
\Big( \sum_{\bf k} \, e^{-i {\bf k} \cdot {\bf R}} \,
H_{mn}^{({\rm rot})}({\bf k}) \Big) / N_{\rm kp} \nonumber \\
&=& \big< w_{m \bf 0} \big| \hat{H} \big| w_{n \bf R} \big>,
\end{eqnarray}
where $\hat{H}$ is the effective one-particle Hamiltonian.
Finally we Fourier transform back to an
arbitrary $k$-point,
\begin{equation}
\label{ftt_2}
H_{mn}^{({\rm rot})}({\bf k'}) = 
\sum_{\bf R} \, e^{i {\bf k'} \cdot {\bf R}} \,
H_{mn}({\bf R}),
\end{equation}
and diagonalize the resulting matrix to find the interpolated 
energy eigenvalues.

\subsection{Inner energy window}
\label{sec:inner_window}

In some situations one wants to construct orbitals that describe the 
original bands {\it exactly} only in a limited energy range.
This can occur when studying transport properties for which
only the states within some small energy range of the Fermi level 
(say, $\pm 1$~eV) 
are relevant.
The challenge is to construct orbitals that achieve that goal while 
remaining as localized as possible. What the resulting interpolated bands look
like outside the energy range of interest is largely immaterial, since it will
not affect the low-energy physics. (Typically they will tend to remain close
in energy to the target range of interest.\cite{andersen00})

A simple extension of the formalism described in the previous sections can 
produce such orbitals.
The idea is to introduce a second (``inner'') energy window --~contained 
within our original (``outer'')
window~-- inside which the original bands are to be described exactly. Let
$M_{\bf k}$ be the number of bands that fall within the inner window at
${\bf k}$, so that $M_{\bf k} \leq N \leq N_{\bf k}$. 
Then we have to minimize $\Omega_{\rm I}$ under the constraint that the
$M_{\bf k}$ original Bloch states inside the inner window must be included in
the subspace ${\cal S}({\bf k})$. We are therefore only free to choose the 
remaining $N-M_{\bf k}$ states when constructing ${\cal S}({\bf k})$. 
Those will have to be extracted from the subspace
spanned by the $N_{\bf k}-M_{\bf k}$ original Bloch eigenstates that are 
inside the outer window but outside the inner window.
That can be achieved by a straightforward modification of the iterative 
procedure
described in Sec.~\ref{sec:min}: The matrix $Z^{(i)}({\bf k})$
in Eq.~(\ref{z_matrix})
becomes an $(N_{\bf k}-M_{\bf k}) \times (N_{\bf k}-M_{\bf k})$ matrix, and
we pick the $N-M_{\bf k}$ leading eigenvectors.

The only remaining issue is how to modify the initialization procedure of
Sec.~\ref{sec:initial_guess} in order to accommodate the inner 
window. Since the first $M_{\bf k}$ basis vectors of the trial subspaces
${\cal S}({\bf k})$ are predetermined, we want the modified procedure to 
provide the
remaining $N-M_{\bf k}$ vectors. Let ${\cal G}({\bf k})$ be an $N$-dimensional space
obtained by projecting the $N$ trial
orbitals onto the $N_{\bf k}$ states inside the outer window,
as described in Sec.~\ref{sec:initial_guess}. Let
$P_{\cal G}({\bf k})$ be the $N_{\bf k} \times N_{\bf k}$ matrix that is the
projection operator onto ${\cal G}({\bf k})$ as expressed in the space
${\cal F}({\bf k})$. Similarly, define $P_{\rm inner}({\bf k})$ as the
$N_{\bf k} \times N_{\bf k}$ projection matrix onto the inner window states, 
and $Q_{\rm inner}({\bf k})={\bf 1}-P_{\rm inner}({\bf k})$. 
Then choose the remaining $N-M_{\bf k}$
basis vectors to be the eigenvectors corresponding to the $N-M_{\bf k}$
largest eigenvalues of 
\begin{equation}
\label{modified_proj}
Q_{\rm inner}({\bf k}) P_{\cal G}({\bf k}) Q_{\rm inner}({\bf k})
| v \rangle = \lambda | v \rangle.
\end{equation}
Such vectors have the desired properties: (i) They are orthogonal
to the states inside the inner window,
and (ii) because
$\lambda= \langle v| P_{\cal G}({\bf k}) | v \rangle$, it is clear that by
choosing the eigenvectors with the largest eigenvalues we guarantee that their
overlap with the space ${\cal G}({\bf k})$ is as large as possible,
while satisfying the constraint (i).

Other kinds of constraints on the minimization of $\Omega_{\rm I}$ may also
be useful. For instance, one might want to ``pin down'' the desired bands at
high-symmetry $k$-points to ensure that the interpolated bands coincide with 
them at those points.

\section{Results}
\label{sec:results}

\subsection{Computational details}
\label{sec:details}

The calculations were performed within the local-density approximation to
density-functional theory, using a plane-wave basis set and Troullier-Martins
norm-conserving pseudopotentials\cite{troullier91} 
in the Kleinman-Bylander representation.
The energy cutoff was set to 75~Ry 
for copper and 35~Ry for silicon, and the
lattice constants were $6.822$~bohr and $10.260$~bohr respectively.
The computed self-consistent Bloch eigenfunctions and eigenvalues that fell
inside the prescribed energy window were stored to disk. They were used as the
input for the minimization of
$\Omega_{\rm I}$, which was carried out as a separate, postprocessing
operation. This produced an optimal subspace characterized by a new set of $N$
Bloch eigenfunctions and eigenvalues
per $k$-point, which were taken as the input for constructing the
``maxloc'' WFs and the interpolated bands. 
In all the cases we have
found the ``maxloc'' WFs to be real (apart from an overall phase factor),
as was already the case when dealing with isolated groups of bands.\cite{mv97}
The self-consistent calculations were performed on a
$10 \times 10 \times 10$ Monkhorst-Pack mesh of $k$-points for copper, and
$6 \times 6 \times 6$ for silicon.
During the minimization of $\Omega_{\rm I}$ and $\widetilde{\Omega}$
a $10 \times 10 \times 10$ uniform grid was used for both copper and silicon.
This grid was shifted in order to include the
$\Gamma$ point ($k=0$), so as to ensure that the ``maxloc'' WFs have the
desired symmetry properties among themselves. 
(For instance, if a grid is used for silicon that does not include $\Gamma$,
the four antibonding WFs in a unit cell do not all have the same spread.)
The mixing parameter $\alpha$ in Eq.~(\ref{linear_mix}) was set to 0.5.

\subsection{Copper}
\label{sec:copper}

\begin{figure}[t]
\centerline{\epsfig{file=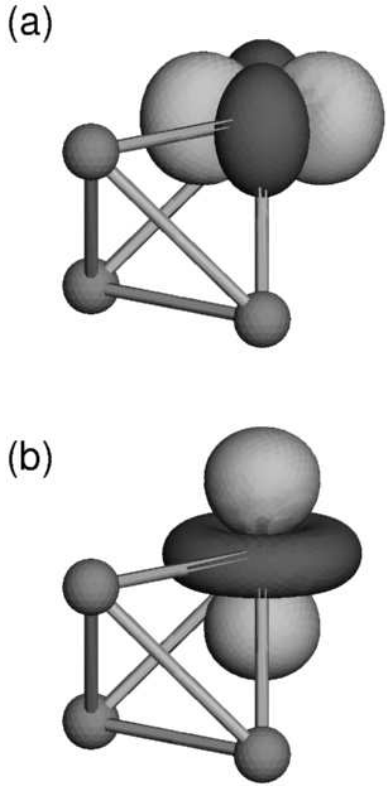,width=2.0in,
angle=0}}
\vspace{0.2cm}
\caption{Contour-surface plots of the two $e_g$ Wannier functions associated
with the ``disentangled'' $d$ bands of copper shown in
Fig.~\ref{fig:Cu_d_bands}(b). The amplitudes are $+0.5/\sqrt{v}$ (light gray)
and $-0.5/\sqrt{v}$ (dark gray), where $v$ is the volume of the primitive cell.
}
\label{fig:Cu_d_wf} 
\end{figure}

Wannier functions for noble and transition metals have previously been computed
using various approaches.\cite{goodings69,modrak87,modrak92,sporkmann94} Below,
taking copper as an example, we show how the present scheme can be used to 
``disentangle'' the narrow $d$ bands from the nearly-free-electron
bands, 
allowing us to treat each group of
WFs separately. Alternatively, one can also treat the 
narrow and the nearly-free-electron bands as a single group.

\subsubsection{Narrow $d$ bands}
\label{sec:copper_narrow}

\begin{table}[b]
\caption{Variation of the optimal Wannier spread $\Omega$ and its 
gauge-invariant part $\Omega_{\rm I}$ (in $\mbox{bohr}^2$) with the choice of 
energy window range (in eV), for the $d$ bands of copper.}
\begin{tabular}{dddd}
\multicolumn{2}{c}{Window range} &
\multicolumn{2}{c}{Total spread} \\
Min & Max & $\Omega_{\rm I}$ & $\Omega$\\
\tableline
$-$9.59 & $-$0.29 & 15.373 & 16.489\\
$-$9.59 & 2.21 & 10.404 & 10.621\\
$-$9.59 & 7.21 & 8.483 & 8.556\\
$-$9.59 & 12.21 & 7.634 & 7.667
\end{tabular}
\label{table:cu_range}
\end{table}

First, an energy window was chosen such that at each $k$-point in the grid it 
contained six or seven energy eigenvalues. As indicated in 
Fig.~\ref{fig:Cu_d_bands}, the precise range of the window is largely
at our disposal; unless explicitly stated otherwise, the numbers given below 
pertain to Fig.~\ref{fig:Cu_d_bands}(b).
In order to extract the five $d$ bands,
we set $N=5$ and initialized the minimization of both $\Omega_{\rm I}$ and
$\widetilde{\Omega}$ from five trial Gaussians of r.m.s. width
1~bohr, each modulated by a different $l=2$ angular eigenfunction. 
After $\sim 50$ iterative steps $\Omega_{\rm I}$ was fully converged, having
decreased from an initial value of 9.957~bohr$^2$ to 8.483~bohr$^2$.
During the subsequent minimization of $\widetilde{\Omega}$ the total Wannier
spread $\Omega$ decreased only slightly, from 8.563~bohr$^2$ to 8.556~bohr$^2$.
In agreement with previous experience on isolated groups of bands,\cite{mv97}
we found for the $d$ bands that at the minimum 
$\Omega_{\rm I} \gg \widetilde{\Omega}$.

The bands obtained by interpolation using the five ``maxloc'' WFs are shown
as dotted lines in Fig.~\ref{fig:Cu_d_bands}, together with the original
band structure. As expected, whenever the dispersive $s$-like
band is far from the narrow
$d$ bands, so that they retain their separate identities, the interpolated 
bands
are very close to the narrow bands. However, whenever the six bands are close
together, and thus strongly hybridized,
the interpolated bands remain narrow, which suggests that
they are mainly $d$-like in character. (Heuristically they can be viewed
as the bands obtained by artificially ``switching off'' the Hamiltonian matrix
elements between $s$ and $d$ WFs, i.e., by removing the hybridization.) 
The $d$ character is confirmed by inspection of the
contour-surface plots of the
``maxloc'' WFs, two of which are shown in Fig.~\ref{fig:Cu_d_wf}.
The quadratic spreads of the five WFs are not exactly 
equal, because of the $e_g \, - \, t_{2g}$ splitting of the $d$-states; 
those shown in Fig.~\ref{fig:Cu_d_wf} ($e_g$ orbitals)
have a spread of 1.700~bohr$^2$ each, whereas the 
remaining three ($t_{2g}$ orbitals) each have a spread of 1.718~bohr$^2$.
These numbers are only slightly larger than the ones reported in Table~III 
of Ref.~\onlinecite{sporkmann94}, obtained using a different method and
a sparser sampling of the Brillouin zone.

\begin{figure}
\centerline{\epsfig{file=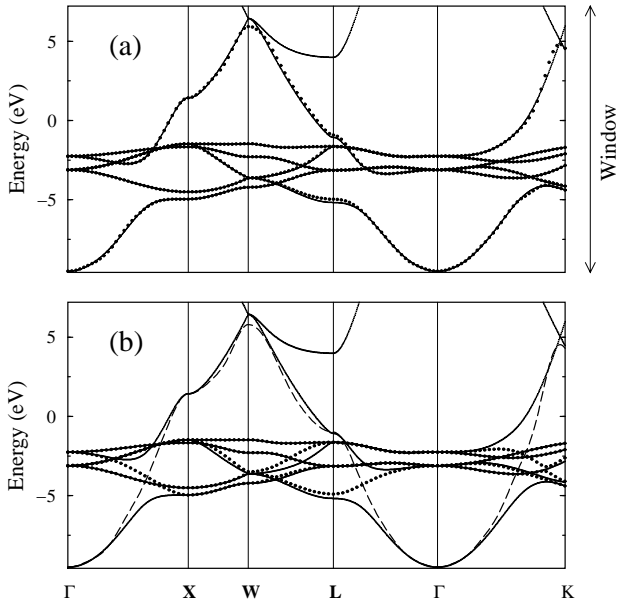,width=3.1in,angle=-90}}
\vspace{0.2cm}
\caption{(a) Dotted lines: the $s$-$d$ bands of copper obtained
by extracting the optimal
six-dimensional subspace ${\cal S}_6({\bf k})$ inside the window.
(b)
Dotted lines: $d$ bands associated with optimal five-dimensional subspace
${\cal S}_5({\bf k})\subset {\cal S}_6({\bf k})$. 
Dashed line: $s$ band ${\cal S}_1({\bf k})$ isolated by taking the
complement of ${\cal S}_5({\bf k})$.}
\label{fig:Cu_s_band}
\end{figure}

In our procedure there is one adjustable parameter, namely the range of the
energy window. This range should be wide enough that it
encompasses the bands of interest, but not be so wide that it also
includes other bands of similar character (e.g., higher $d$ bands). In
the limit
of a very wide window the spaces ${\cal F}({\bf k})$ would
contain a complete set of states, so that by mixing in states far away from the
energy range of interest but of similar character, the
spread of the WFs could be made arbitrarily small (and the corresponding bands
would become flat). 
Table~\ref{table:cu_range}
shows how the optimal Wannier spreads are affected by varying the window
range within reasonable bounds. As anticipated,
the spread decreases with increasing energy range.\cite{foot:range}
The change in the interpolated energy bands is less pronounced, although they 
do become somewhat narrower
(compare Figs.~\ref{fig:Cu_d_bands}(a) and \ref{fig:Cu_d_bands}(b)).
In particular, the upward shift of the lowest interpolated band at $L$ is 
caused by mixing with the seventh band, which has the same symmetry label 
($L_1$).\cite{burdick63}

\subsubsection{Nearly-free-electron band}
\label{sec:copper_wide}
 
The unconstrained minimization of $\Omega_{\rm I}$ usually produces narrow
bands, since the character of the Bloch states in
such bands tends to have only a small variation
across the Brillouin zone, corresponding to well-localized electrons 
(this may not be the case in the presence of avoided crossings).
The method is therefore ideally suited for
directly extracting the narrow $d$ bands from the $s$-$d$ complex. If instead
one is interested in isolating the wider, nearly-free-electron $s$ band,
direct minimization of $\Omega_{\rm I}$ for one-dimensional subspaces
is not the appropriate strategy. Instead one can proceed as follows.
First choose an energy window that includes the $s$-$d$ band complex
(we used the one indicated in Fig.~\ref{fig:Cu_d_bands}(b)).
Then minimize $\Omega_{\rm I}$
with $N=6$; this produces a six-dimensional subspace ${\cal S}_6({\bf k})$
throughout the Brillouin zone that consists of the $s$--$d$ band complex. 
Next extract the five $d$ bands
by minimizing $\Omega_{\rm I}$ within ${\cal S}_6({\bf k})$ choosing
$N=5$; this yields a 
space ${\cal S}_5({\bf k}) \subset {\cal S}_6({\bf k})$. The difference
between the two is a one-dimensional space ${\cal S}_1({\bf k})$
containing the desired band.
Fig.~\ref{fig:Cu_s_band}(a) shows the bands associated with
${\cal S}_6({\bf k})$, and Fig.~\ref{fig:Cu_s_band}(b) shows the bands 
corresponding to ${\cal S}_5({\bf k})$ and ${\cal S}_1({\bf k})$.

\begin{table}[t]
\caption{Spreads of the ``maxloc'' WFs for the
separate $d$-band and $s$-band subspaces (${\cal S}_5$ and 
${\cal S}_1$), and for the combined $s$-$d$ subspace
${\cal S}_6$. The numbers in 
parentheses are the $\Omega_{\rm I}$ values, and $t$ stands\
for tetrahedral-interstitial-centered orbital.
The corresponding bands are displayed in Fig.~\ref{fig:Cu_s_band}.}
\begin{tabular}{cdd|cdd}
\multicolumn{3}{c}{Two separate subspaces} &
\multicolumn{3}{c}{One combined subspace} \\
\tableline
$d_{e_g   }$ & 1.710 & & $d_{e_g   }$ & 1.731 & \\
$d_{e_g   }$ & 1.710 & & $d_{e_g   }$ & 1.731 & \\
$d_{t_{2g}}$ & 1.808 & & $d_{t_{2g}}$ & 2.328 & \\
$d_{t_{2g}}$ & 1.808 & & $d_{t_{2g}}$ & 2.328 & \\
$d_{t_{2g}}$ & 1.808 & & $d_{t_{2g}}$ & 2.254 & \\
$\Omega_{\rm min}[{\cal S}_5]$ & 8.844 & (8.745) & & & \\
$t$          &12.929 & & $t$          &10.263 & \\
$\Omega_{\rm min}[{\cal S}_1]$ & 12.929 & (10.826) &
$\Omega_{\rm min}[{\cal S}_6]$ & 20.634 & (16.506)
\end{tabular}
\label{table:cu_subspaces}
\end{table}

\begin{figure}
\centerline{\epsfig{file=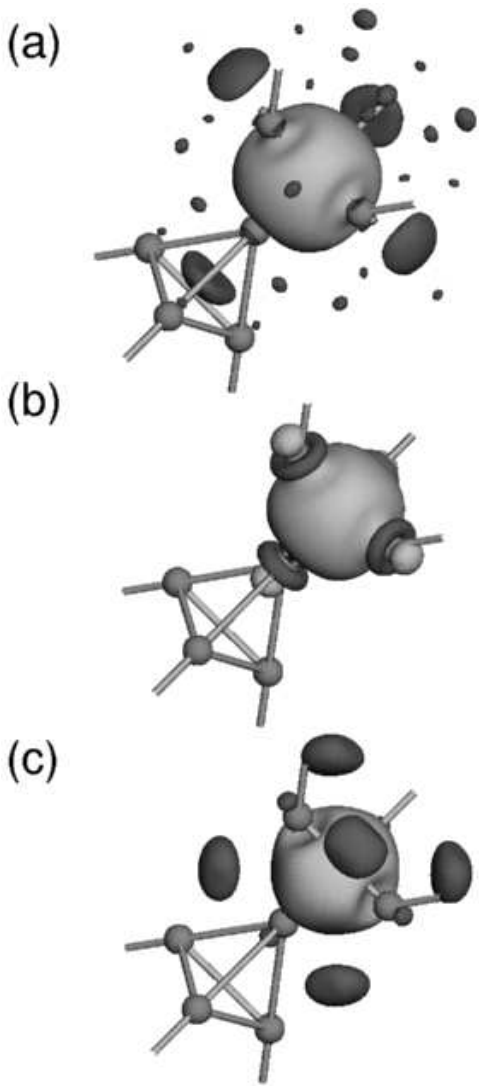,width=2.5in,angle=0}}
\vspace{0.2cm}
\caption{Contour-surface plots of interstitial-centered
``maxloc'' WFs. (a) $t$-like WF associated with the subspace 
${\cal S}_6({\bf k})$
of Fig.~\ref{fig:Cu_s_band} and Table~\ref{table:cu_subspaces};
(b) WF associated with the band in Fig.~\ref{fig:Cu_frozen_band};
(c) $t$-like WF associated with the subspace ${\cal S}_7({\bf k})$
in Fig.~\ref{fig:Cu_t2d5_bands}(a) and Table~\ref{table:cu_subspaces_b}.
The amplitudes are $+0.5/\sqrt{v}$ (light gray) and
$-0.17/\sqrt{v}$, $-0.3/\sqrt{v}$, and $-0.25/\sqrt{v}$ (dark gray) in
(a), (b), and (c) respectively.
}
\label{fig:blobs}
\end{figure}

In Table~\ref{table:cu_subspaces} are presented the optimal Wannier
spreads for the different subspaces.
We find that the spread of the $s$-like WF 
is considerably smaller than the 
$\sim$45\,bohr$^2$ reported in Table~III of Ref.~\onlinecite{sporkmann94}.
Moreover, contrary to what one might
have expected, that WF is centered not on an atom, but on a tetrahedral 
interstitial site, as shown in Fig.~\ref{fig:blobs}(a). 
Since there are two such sites per atom,
a breaking of symmetry must have occurred when selecting the subspace 
${\cal S}_6({\bf k})$. Indeed there are two
degenerate minima of $\Omega_{\rm I}$ with $N=6$, 
one for each of the interstitial sites. If the minimization
is initialized by projecting five $d$-like
orbitals plus one $s$-like orbital, all atom-centered, the breaking of
symmetry occurs spontaneously during the iterative procedure
(the minimization of $\Omega_{\rm I}$ reaches a plateau, presumably a
saddle point, and eventually the algorithm finds its way towards one of the two
minima). If instead the $s$ trial orbital is centered around one of the 
tetrahedral interstitial sites, the minimization starts inside
the basin of attraction of the corresponding minimum.

\begin{figure}
\centerline{\epsfig{file=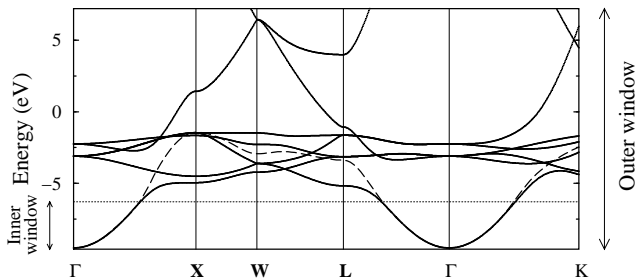,width=1.4in,angle=-90}}
\vspace{0.2cm}
\caption{Dashed line: Band obtained using both an inner and an outer
energy window.
}
\label{fig:Cu_frozen_band}
\end{figure}
Finally, as a simple illustration of the ``inner window'' idea of
Sec.~\ref{sec:inner_window}, we show in
Fig.~\ref{fig:Cu_frozen_band} the single band ($N$=1)
that results when an inner window is selected in the energy range below the
$d$ bands. As expected, the interpolated band is identical to the original one
inside that window.
Moreover, it remains quite narrow outside, where it acquires a
pronounced $d$ character. (This means that the cost in
$\Omega_{\rm I}$ of changing from $s$ to $d$ character is more than
compensated by the smaller dispersion --~and hence smaller $\Omega_{\rm I}$~--
of the more localized $d$-like states.)
Accordingly, the ``maxloc'' WF, shown in Fig.~\ref{fig:blobs}(b),
is again centered at a tetrahedral interstitial site, like the
WF of Fig.~\ref{fig:blobs}(a),
but now it has a substantial admixture of
$d$-like satellites and a smaller spread, $\Omega=7.323$~bohr$^2$
($\Omega_{\rm I}=7.306$~bohr$^2$).

The results of this Section indicate that
the occurrence of a symmetry breaking in the minimization of $\Omega_{\rm I}$
with a ``maxloc'' WF centered at a tetrahedral interstitial site appears to be
a rather robust result.

\subsubsection{Symmetric two-WF description of dispersive bands}
\label{sec:copper_blobs}

\begin{figure}
\centerline{\epsfig{file=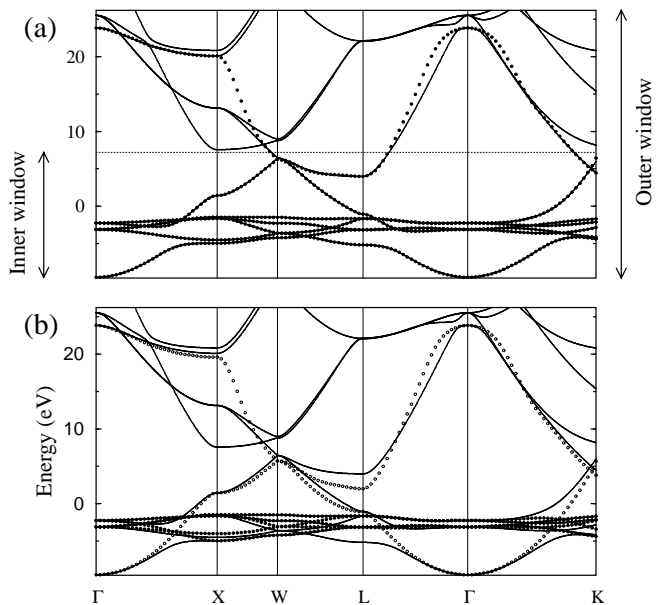,width=3.1in,angle=-90}}
\vspace{0.2cm}
\caption{(a) Dotted lines: Interpolated bands associated with the optimal
subspace ${\cal S}_7({\bf k})$ containing five $d$-like WFs and
two tetrahedral-interstitial-centered WFs. (b) Dark dotted lines:
$d$ bands associated with optimal five-dimensional subspace
${\cal S}_5^{'}({\bf k})\subset {\cal S}_7({\bf k})$.
Light dotted lines: dispersive bands $S_2({\bf k})$ isolated by taking the
complement of ${\cal S}_5^{'}({\bf k})$.
}
\label{fig:Cu_t2d5_bands}
\end{figure}

Remarkably, we find that the symmetry can be restored, and a more
faithful overall description of the bands can be achieved, by bringing in
just one more dispersive band and working with a {\it set of seven WFs}.
More precisely, we choose an energy window such as the
one indicated in Fig.~\ref{fig:Cu_t2d5_bands}(a), containing seven or more 
bands, and minimize $\Omega_{\rm I}$ with $N=7$. 
(To ensure that the low-energy part of the
band complex is well described, we freeze it inside an inner window.) After 
applying the localization procedure we obtain, besides the five $d$ orbitals, 
{\it two equivalent WFs, each centered at one of the two tetrahedral
interstitial sites}.  One of the latter is shown in Fig.~\ref{fig:blobs}(c). 
The optimal Wannier spreads are given in Table~\ref{table:cu_subspaces_b};
it can be seen that the spread of each of the two interstitial WFs is
considerably smaller than that of the single interstitial WF in
Table~\ref{table:cu_subspaces} and Fig.~\ref{fig:blobs}(a).

\begin{table}
\caption{Spreads of the ``maxloc'' WFs for the
separate $d$-band and low-lying dispersive bands subspaces 
(${\cal S}_5^{'}$ and ${\cal S}_2$), and for the combined subspace
${\cal S}_7$. The numbers in 
parentheses are the $\Omega_{\rm I}$ values, and $t$ stands\
for tetrahedral-interstitial-centered orbital.
The corresponding bands are displayed in Fig.~\ref{fig:Cu_t2d5_bands}.}
\begin{tabular}{cdd|cdd}
\multicolumn{3}{c}{Two separate subspaces} &
\multicolumn{3}{c}{One combined subspace} \\
\tableline
$d_{e_g   }$ & 1.687 &                  & $d_{e_g   }$ & 1.687 & \\
$d_{e_g   }$ & 1.686 &                  & $d_{e_g   }$ & 1.687 & \\
$d_{t_{2g}}$ & 1.472 &                  & $d_{t_{2g}}$ & 1.737 & \\
$d_{t_{2g}}$ & 1.472 &                  & $d_{t_{2g}}$ & 1.737 & \\
$d_{t_{2g}}$ & 1.472 &                  & $d_{t_{2g}}$ & 1.737 & \\
$\Omega_{\rm min}[{\cal S}_5^{'}]$ & 7.788 & (7.751) & & & \\
$t$          & 8.568 &                  & $t$          & 7.812 & \\
$t$          & 8.568 &                  & $t$          & 7.812 & \\
$\Omega_{\rm min}[{\cal S}_2]$ & 17.136 & (16.822) &
$\Omega_{\rm min}[{\cal S}_7]$ & 24.209 & (22.034)
\end{tabular}
\label{table:cu_subspaces_b}
\end{table}

Fig.~\ref{fig:Cu_t2d5_bands}(b) shows the $d$-like bands associated with the
optimal five-dimensional subspace
${\cal S}_5^{'}({\bf k})\subset {\cal S}_7({\bf k})$, as well as the
dispersive bands associated with $S_2({\bf k})$, the complement of
${\cal S}_5^{'}({\bf k})$ inside ${\cal S}_7({\bf k})$.
There is an upward shift in energy of the states $X_{3}$, $W_{3}$, and
$L_{1}$ in the narrow bands, due to mixing with the states of the same 
symmetry in the dispersive bands, which suffer a downward shift
of the same magnitude.

The fact that our procedure naturally generates a pair of WFs centered
at the tetrahedral interstitial sites can be rationalized in terms of a 
tight-binding description of the nearly-free electron states.
The tetrahedral interstitial sites form a simple cubic lattice, so
that in view of Fig.~\ref{fig:blobs}(c) one might imagine that
the electronic states of these WFs would be roughly analogous to those
of a nearest-neighbor tight-binding model of $s$ orbitals on the sites
of a simple cubic lattice.  Indeed we have checked that the main
qualitative features of the interpolated bands associated with the
two interstitial-centered WFs (light dotted lines in 
Fig.~\ref{fig:Cu_t2d5_bands}(b)) are captured by such a
tight-binding model, but folded back into the fcc Brillouin zone to
give two bands instead of one.

\begin{table}[t]
\caption{A list, in order of increasing energy, of the symmetry labels of 
selected states in the band structure of copper
(taken from Ref.~\protect\onlinecite{burdick63}), and whether or not they are
captured by each of the tight-binding bases discussed in the text. An asterisk
$(^{*})$ indicates that the state is occupied.}
\begin{tabular}{llccc}
                  & Degeneracy & $sd^5$   & $t^2d^5$ & $sp^3d^5$\\
\hline
$\Gamma_1$        & \qquad 1$^{*}$    & yes      & yes      & yes\\
$\Gamma_{25'}$    & \qquad 3$^{*}$    & yes      & yes      & yes\\
$\Gamma_{12}$     & \qquad 2$^{*}$    & yes      & yes      & yes\\
$\Gamma_{2'}$     & \qquad 1          & --       & yes      & -- \\
$\Gamma_{15}$     & \qquad 3          & --       & --       & yes\\
\hline
$X_1$             & \qquad 1$^{*}$    & yes      & yes      & yes\\
$X_3$             & \qquad 1$^{*}$    & yes      & yes      & yes\\
$X_2$             & \qquad 1$^{*}$    & yes      & yes      & yes\\
$X_5$             & \qquad 2$^{*}$    & yes      & yes      & yes\\
$X_{4'}$          & \qquad 1          & --       & yes      & yes\\
$X_1$             & \qquad 1          & yes      & --       & yes\\
$X_{5'}$          & \qquad 2          & --       & --       & yes\\
$X_3$             & \qquad 1          & --       & yes      & -- \\
\hline
$L_1$             & \qquad 1$^{*}$    & yes      & yes      & yes\\
$L_3$             & \qquad 2$^{*}$    & yes      & yes      & yes\\
$L_3$             & \qquad 2$^{*}$    & yes      & yes      & yes\\
$L_{2'}$          & \qquad 1$^{*}$    & --       & yes      & yes\\
$L_1$             & \qquad 1          & yes      & yes      & yes\\
$L_{2'}$          & \qquad 1          & --       & --       & -- \\
$L_{3'}$          & \qquad 2          & --       & --       & yes\\
\hline
$W_{2'}$          & \qquad 1$^{*}$    & yes      & yes      & yes\\
$W_3$             & \qquad 2$^{*}$    & yes      & yes      & yes\\
$W_1$             & \qquad 1$^{*}$    & yes      & yes      & yes\\
$W_{1'}$          & \qquad 1$^{*}$    & yes      & yes      & yes\\
$W_3$             & \qquad 2          & --       & yes      & yes\\
$W_{2'}$          & \qquad 1          & --       & --       & yes\\
$W_1$             & \qquad 1          & yes      & --       & yes\\
\end{tabular}
\label{table:symm}
\end{table}
The quality of the interpolated bands in Fig.~\ref{fig:Cu_t2d5_bands}(a)
suggests that the two tetrahedral-interstitial-centered orbitals 
(which we denote as $t$ orbitals) complement the five atom-based $d$ 
orbitals nicely to form a basis ($t^2d^5$) for a tight-binding
parametrization of the copper bands. This requires only one more basis
function than the traditional ``minimal basis''\cite{harrison80} $sd^5$ 
(five $d$ plus one $s$ atomic orbitals), while still remaining
more economical than the $sp^3d^5$ basis.\cite{papaconstantopoulos86}
The three bases are compared in Table~\ref{table:symm}. At each 
high-symmetry $k$-point we list, in order of increasing energy, the 
symmetry labels of the states that occur in a detailed band-structure 
calculation (e.g., Ref.~\onlinecite{burdick63}), and then
whether or not they are captured by each of the tight-binding bases.
Inspection of the table clarifies that the $t^2d^5$ basis has some
very attractive features. Whereas the $sd^5$ basis misses the 
$X_{4'}$ state\cite{harrison80} (unoccupied $p$-like state not far
above $E_F$) and, even more importantly, the $L_{2'}$ state (occupied
$p$-like state just below $E_F$),
$t^2d^5$ gets the symmetries right up to at least the first state above
$E_F$ at each high-symmetry $k$-point. 
Even $sp^3d^5$ does not do this, failing at the
$\Gamma$ point, since the state $\Gamma_{2'}$ has $f$ character. 
A consequence of this analysis is that the $t$ orbitals cannot be
constructed solely from $s$ and $p$ orbitals. This can also be seen from 
Fig.~\ref{fig:blobs}(c): 
The positive-amplitude central portion of the WF can be interpreted in terms of
a superposition of four $sp$ hybrids coming from each of the four surrounding 
copper atoms and pointing towards the interstitial; however this picture 
cannot account for the six negative lobes.

To conclude, we note that the $sp^3d^5$
description can also be obtained from our procedure, by minimizing 
$\Omega_{\rm I}$ with $N=9$ 
within a window containing eleven or more bands (e.g., with the upper bound
at 32.2~eV). The ``maxloc'' WFs are then five atom-centered $d$-like orbitals 
plus four equivalent $sp^3$-like hybrids centered near the atom.

\subsection{Silicon}
\label{sec:silicon}

Several authors have previously discussed and computed WFs for silicon and
other tetrahedral semiconductors. Some works have focused on the WFs
associated with the valence 
bands,\cite{mv97,teichler71,kane78,tejedor79,satpathy88,fernandez97} while
others have also dealt with the lowest four conduction 
bands.\cite{sporkmann97,kohn73}

\subsubsection{Bond orbitals}
\label{sec:silicon_bond}

\begin{figure}
\centerline{\epsfig{file=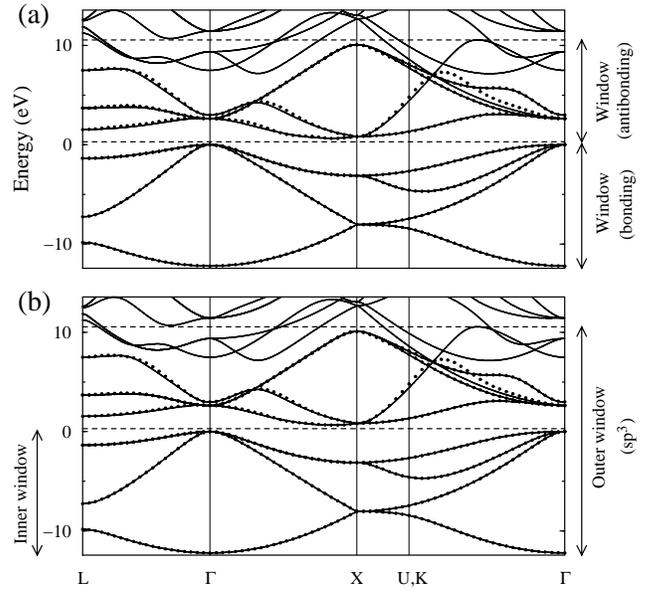,width=3.0in,angle=-90}}
\vspace{0.2cm}
\caption{
Solid lines: Original band structure of silicon.
Dotted lines: Wannier-interpolated bands. In (a) the
valence and low-lying conduction bands are treated separately,
which produces four bonding and four antibonding Wannier functions; 
in (b) they are treated as a single 
group, which yields eight $sp^3$-type Wannier functions.
}
\label{fig:si_bands}
\end{figure}

A set of eight bond-centered WFs, four bonding and four antibonding, can be 
obtained by using separate energy windows for each of the
two groups, as indicated in
Fig.~\ref{fig:si_bands}(a). Since the valence bands form an isolated group,
inside the corresponding window $N_{\bf k}=N=4$ throughout the 
Brillouin zone. Hence
there is no freedom for minimizing $\Omega_{\rm I}$, and one can proceed
directly with the minimization of $\widetilde{\Omega}$ to compute the 
``maxloc'' WFs, as done in Ref.~\onlinecite{mv97}. The resulting
bands are essentially indistinguishable from the original ones, since 
for such a dense $k$-mesh the
interpolation error is very small. The trial orbitals used to start the
minimization were bond-centered Gaussians with a root mean-square (r.m.s.)
width of 1.89~bohr. The value of the optimal spread was 
$\Omega=30.13$~bohr$^2$, of which 28.39~bohr$^2$ came from $\Omega_{\rm I}$.

The use of an energy window becomes necessary for the four low-lying empty 
bands, which are attached to higher bands. As trial orbitals we used
an antibonding combination of Gaussians with a
r.m.s. width of 1~bohr. Each Gaussian was sitting halfway between one of the 
two atoms 
and the center of their common bond. During the minimization $\Omega_{\rm I}$ 
decreased
from $106.76$~bohr$^2$ to $87.47$~bohr$^2$, having reached the minimum in
less than 30 steps.
(An alternative is to choose the initial subspace at each ${\bf k}$ as the 
lowest four energy eigenstates inside the energy window. This yields an initial
$\Omega_{\rm I}=98.10$~bohr$^2$, and again the absolute minimum is reached
after $\sim$30 steps.)
The total spread of the four ``maxloc'' WFs was
$\Omega=97.49$~bohr$^2$; as expected,\cite{teichler71} this
is considerably larger than for the bonding WFs. Note also that
$\widetilde{\Omega}$ accounts for more than $10\%$ of the total spread, whereas
for the bonding ``maxloc'' WFs that number was less than 6\%.
This is related to the fact that the antibonding WFs are more spread
out, causing matrix elements of the type
$\left< w_{m \bf R} | {\bf r} | w_{n \bf 0} \right>$ with 
${\bf R} \not= {\bf 0}$  to have larger values. Eq.~(15) of 
Ref.~\onlinecite{mv97} 
shows that this results in a larger $\widetilde{\Omega}$. The very small 
contribution of $\widetilde{\Omega}$ to the total spread of the highly 
localized $d$-like WFs in copper (less than $1\%$), as well as the
comparatively larger contribution in the interstitial-centered WFs are thus 
easily understood.

\begin{figure}[t]
\centerline{\epsfig{file=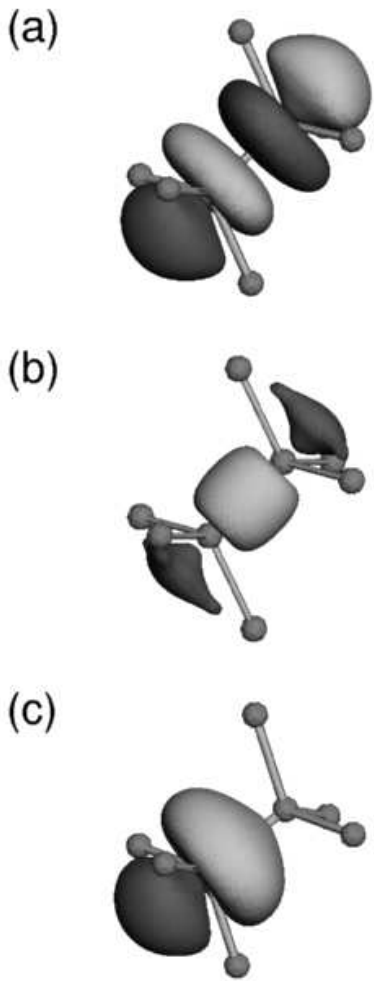,width=2.0in,angle=0}}
\vspace{0.2cm}
\caption{Contour-surface plots of Wannier functions in silicon.
(a) Antibonding, (b) bonding, and
(c) $sp^3$-type. In (a) and (c) the amplitudes are $+0.5/\sqrt{v}$ (light gray)
and $-0.5/\sqrt{v}$ (dark gray); in (b) they are $+1.4/\sqrt{v}$ (light gray)
and $-0.4/\sqrt{v}$ (dark gray).
}
\label{fig:si_wf} 
\end{figure}

In Fig.~\ref{fig:si_wf}(a) we present the contour-surface plot of one
``maxloc'' antibonding WF in silicon. The other three are identical (related to
the first by the tetrahedral symmetry operations). Fig.~\ref{fig:si_wf}(b) 
shows one of the four identical bonding WFs.

\subsubsection{$sp^3$ hybrids}
\label{sec:silicon_sp3}

As discussed in Ref.~\onlinecite{kohn73}, one may instead treat the 
four valence and four low-lying conduction bands as a
single group, which leads to ``maxloc'' WFs of $sp^3$ character 
(Fig.~\ref{fig:si_wf}(c)). Using our method this may be done as indicated in 
Fig.~\ref{fig:si_bands}(b).
An outer energy window is chosen which spans the eight bands of interest, and
the valence bands are ``frozen'' inside an inner window; this ensures that
they are not affected by the minimization of $\Omega_{\rm I}$, whose only
aim is to extract the four low-lying antibonding bands from the conduction band
complex. We have started the minimization of $\Omega_{\rm I}$ in two different
ways: (i) by projecting eight ``atom-centered'' $sp^3$-type combinations of 
Gaussians, and (ii) by projecting four bond-centered Gaussians plus four 
antibonding combinations of Gaussians, as done in the previous Section. 
In both cases the 
minimization took about $20$ steps, taking from 76.04~bohr$^2$ in the former
case and 84.08~bohr$^2$ in the latter to 63.50~bohr$^2$. 
As for the minimization of
$\widetilde{\Omega}$, the absolute minimum 
($\Omega=85.41$~bohr$^2$)  was reached only with (i); with
(ii) the algorithm became trapped in a local minimum 
($\Omega=101.97$~bohr$^2$)
having the same symmetry
as the trial orbitals, with four bonding (antibonding) WFs with a spread of
$6.37$~bohr$^2$ ($19.12$~bohr$^2$) each.

We end this section with the following observation.
Suppose we take the four-dimensional
valence (bonding) space ${\cal S}_4^{(b)}({\bf k})$
together with the optimal four-dimensional antibonding subspace
${\cal S}_4^{(a)}({\bf k})$
(Fig.~\ref{fig:si_bands}(a)) to form an eight-dimensional space
${\cal S}^{'}_8({\bf k})={\cal S}_4^{(b)}({\bf k}) \cup
{\cal S}_4^{(a)}({\bf k})$. 
This space has $\Omega_{\rm I}=63.64$~bohr$^2$, which is slightly
higher than the value $63.50$~bohr$^2$ associated with the optimal subspace 
${\cal S}_8({\bf k})$ for the eight-band problem with an inner
window (Fig.~\ref{fig:si_bands}(b)). Thus, if we take 
${\cal S}^{'}_8({\bf k})$ as an initial guess for the minimization of
$\Omega_{\rm I}$ in the eight-band problem with an inner window, we will be 
starting slightly above
the absolute minimum. 
The extra reduction in $\Omega_{\rm I}$ comes about
because the functional that is minimized to obtain
${\cal S}_8({\bf k})$ contains terms involving overlap between
low-lying conduction states at ${\bf k}$ and valence states at 
neighboring ${\bf k}+{\bf b}$.  The wavefunctions relax in response
to these extra terms, and consequently the two antibonding
subspaces are not exactly the same.  However, they are almost identical, and
therefore the same is true for the interpolated bands
(compare Figs.~\ref{fig:si_bands}(a) and \ref{fig:si_bands}(b)).

\section{Conclusions}
\label{sec:conclusions}

We have discussed and implemented a practical method for extracting
maximally-localized Wannier functions from entangled energy bands, starting 
from the Bloch eigenfunctions obtained in a standard electronic structure 
calculation. 
Our method is based on a prescription for ``disentangling'' the bands of 
interest from the rest of
the band complex inside an energy window specified by the user. The idea 
is to extract a subspace of Bloch-like states
whose character varies as little and as smoothly as possible across the 
Brillouin zone.
This is achieved by minimizing a functional which measures the ``spillage'',
or change of character of the subspace across the Brillouin zone.
The present scheme can be viewed as an extension of the maximally-localized
Wannier function method of Marzari and Vanderbilt,\cite{mv97} which was 
designed to deal
with isolated groups of bands only. More precisely, it introduces an extra step
--~the construction of the optimal subspace~-- which is followed by the
determination of the ``maxloc'' WFs by applying the 
localization algorithm of Marzari and Vanderbilt to that subspace. 
The procedure for determining this optimal subspace is both
stable and computationally very fast.

Some possible applications of such WFs have been mentioned in the Introduction.
Of particular interest is the ability to
obtain WFs for the low-lying empty or partially filled bands. For instance, it
has been suggested that these could be useful for accurate calculations of the 
optical properties of semiconducting nanocrystals.\cite{mizel97}
Another potential use
of the present method could arise in
the description of surface states (e.g., Ref.\onlinecite{hellberg99}), in 
particular when the surface bands become resonant with the bulk bands. 
The striking result that we have obtained for the low-lying 
broad bands of copper, with the WFs being centered at the
tetrahedral interstitial sites, suggests that the method may provide 
insight into the chemistry of transition metal compounds. 
Also, since the
``maxloc'' WFs provide a compact interpolation scheme for the band 
structure, they could be used as part of an efficient algorithm for determining
the Fermi surface.
Finally, it might be interesting to apply the present ideas to the 
construction of lattice WFs describing the part
of the phonon spectrum relevant for studying structural phase 
transitions.\cite{rabe95,iniguez00}

\section*{Acknowledgments}

This work was supported by the NSF Grant DMR-9981193. We would like to thank 
Dr. Noam Bernstein for providing us with his visualization software {\it dan}.


\begin{references}
\bibitem{foot:omega_i_mix} For $\alpha \not=1$ Eqs.~\ref{omega_i_k} and 
\ref{omega_i_k_b} are not equivalent, except at self-consistency.
\bibitem{foot:one_band}
For a single band the interpolated band structure is gauge-invariant,
so that the smoothness is independent of the degree of localization of the
WF.
\end{references}
\end{document}